\documentclass[
numbers=noenddot,
twoside,12pt,a4paper,
abstract=true
]{scrartcl}

\usepackage[latin1]{inputenc}          
\usepackage[T1]{fontenc}
\usepackage[final,babel=true,expansion=true,protrusion=true,spacing=true,kerning=true]{microtype}
\usepackage[adobe-utopia,expert]{mathdesign} 
\usepackage[ttscale=.875]{libertine}

\usepackage{graphicx}                
\usepackage{hyperref}
\hypersetup{linktoc=all,colorlinks=true,urlcolor=red!80,linkcolor=green!70!blue!100!,citecolor=green!70!blue!100!}
\usepackage{amsthm}
\usepackage[framemethod=tikz]{mdframed}
\usepackage{pgfplots} 
\pgfplotsset{compat=1.11}
\usepackage{tikz}
\usepackage{enumitem}
\usepackage{bbm} 
\usepackage{lastpage}
\usepackage{makeidx}
\usepackage[nowatermark]{fixmetodonotes}
\usepackage{epstopdf}
\usepackage{epsfig}
\usepackage[nonumberlist,nopostdot]{glossaries} 
\usepackage{mathtools} 
\usepackage[draft=false,headsepline=true,footsepline=false]{scrlayer-scrpage} 
\usepackage[top=2.5cm,left=3cm,right=3cm,bottom=2.5cm]{geometry}

\usepackage{sidecap}
\sidecaptionvpos{figure}{b}
\usepackage{caption}

\usepackage{mymacros}
\graphicspath{{img/}}
 
\mdfdefinestyle{mystyle}{backgroundcolor=lightgray!0,hidealllines=true,leftmargin=0pt,rightmargin=0pt,innerleftmargin=0pt}

{
	\newtheoremstyle{break}{3pt}{3pt}{\normalfont}{}{\bfseries}{}{\newline}{}
	\theoremstyle{break}
	\newmdtheoremenv[style=mystyle]{thm}{Theorem}[section]
	\newmdtheoremenv[style=mystyle]{thm*}{Theorem}
	\newmdtheoremenv[style=mystyle]{cor}[thm]{Corollary}
	\newmdtheoremenv[style=mystyle]{lem}[thm]{Lemma}
	\newmdtheoremenv[style=mystyle]{prop}[thm]{Proposition}
	\newmdtheoremenv[style=mystyle]{dfn}[thm]{Definition}
	\theoremstyle{plain}
	
	\newtheorem*{ex*}{Example}
	\theoremstyle{remark}
	\newtheorem{remark}[thm]{Remark}
	\newtheorem*{rem*}{Remark}
	
}

\newenvironment{bew*}{\vspace{-1.5ex}}{\hspace{1cm}\hspace*{\fill}$\Box$\\%
}

\fontfamily{phv}\selectfont                    
\let\emph\textbf                               

\numberwithin{equation}{section}

\newglossarystyle{mysuper}{
	\glossarystyle{super}
	
	\setlength{\glsdescwidth}{0.8\textwidth}
	}

\automark[section]{subsection}	
\clearscrheadfoot 

\lohead{\leftmark}
\rehead{\rightmark}
\ohead{\pagemark}

\KOMAoptions{headings=small}
\addtokomafont{section}{\centering\normalfont\bfseries\large}
\addtokomafont{subsection}{\centering\normalfont\bfseries\large}
\allowdisplaybreaks[1]  

\usepackage{setspace}
\providecommand{\keywords}[1]{\textbf{Key words:} #1}
\providecommand{\AMS}[1]{\textbf{AMS subject classifications:} #1}

\KOMAoptions{%
    headsepline=false}

\clearscrheadfoot
\cohead{A Phase Field Approach to Compressible Droplet Impingement}
\cehead{Lukas Ostrowski and Christian Rohde}
\ohead{\pagemark}

\author{Lukas Ostrowski\thanks{Institute of Applied Analysis and Numerical Simulation, University of Stuttgart 
  (\href{mailto:Lukas.Ostrowski@mathematik.uni-stuttgart.de}{Lukas.Ostrowski@mathematik.uni-stuttgart.de).}}
\and Christian Rohde\thanks{Institute of Applied Analysis and Numerical Simulation, University of Stuttgart,  (\href{mailto:Christian.Rohde@mathematik.uni-stuttgart.de}{Christian.Rohde@mathematik.uni-stuttgart.de}).}}

\title{A Phase Field Approach to Compressible Droplet Impingement}
\date{}

\begin{document}
\maketitle

\begin{abstract}
We consider the impingement of a droplet onto a wall with high impact speed.  
To model this process we favour a diffuse-interface concept. Precisely, we suggest a compressible Navier--Stokes--Allen--Cahn model following \cite{Dreyer2014}.
Basic properties of the model are discussed. To cope with the fluid-wall interaction, we derive thermodynamically consistent boundary conditions that account for dynamic contact angles. We briefly discuss an discontinuous Galerkin scheme which approximates the energy dissipation of the system exactly and illustrate the results with a series of numerical simulations. Currently, these simulations are restricted to static contact angle boundary conditions.
\end{abstract}

\begin{center}
\parbox{0.9\linewidth}{

\AMS{76T99,65M60}}

\end{center}

\section{Introduction}
\label{sec:intro}

In many fluid dynamic scenarios the compressibility of a liquid is negligible. This allows for simplifications such that direct numerical simulations can rely on simpler incompressible models. In the context of droplet impingement incompressibility is only justified for small impact speeds. High impact speeds trigger compressibility effects of the liquid droplet which can determine the flow dynamics significantly. Examples for high speed droplet impact scenarios can be found in many industrial applications such as liquid-fueled engines, spray cooling or spray cleaning. In \cite{Haller2002} it has been shown that incompressible models are not adequate to describe high speed impacts, especially due to the fact that the jetting dynamics are influenced by a developing shock wave in the liquid phase \cite{Haller2003}. The time after the impact of the droplet until jetting occurs is actually smaller than the predicted time of incompressible models due to the shock wave pattern. In \cite{Haller2002} a compressible sharp-interface model is used for the simulations. However, sharp-interface models become intricate in the presence of changes in droplet topology and contact line motion. For this reason, we introduce a diffuse-interface model in this contribution, namely a compressible Navier--Stokes--Allen--Cahn phase field model which allows for complex interface morphologies and dynamic contact angles.
\section{Phase Field Models}
\label{sec:model}
Phase field models form a special class of diffuse-interface models. In contrast to sharp-interface models, the interface has a (small) finite thickness and in the interfacial region the different fluids are allowed to mix. An additional variable, the \emph{phase field}, is introduced which allows to distinguish the different phases. This concept has the advantage that only one system of partial differential equations on the entire considered domain needs to be solved, whereas for sharp-interface models bulk systems need to be solved which are coupled across the interface by possibly complex conditions. Based on energy principles phase field models can be derived in a thermodynamic framework, see \cite{Anderson1998,Freistuehler2016} for an overview. They
fulfill the second law of thermodynamics meaning that the Clausius--Duhem inequality \cite{Truesdell1952} is fulfilled. In the case of isothermal models this is equivalent to an energy inequality. There are several (quasi-)incompressible \cite{Lowengrub1998,Abels2012}, compressible \cite{Blesgen1999,Dreyer2014,Witterstein2010} and recently even incompressible--compressible phase field models \cite{Repossi2017,Ostrowski2019}. In this section we introduce a compressible Navier--Stokes--Allen--Cahn model.

\subsection{A Compressible Navier--Stokes--Allen--Cahn system}
\label{sec:NSAC}

We consider a viscous fluid at constant temperature. The fluid is assumed to exist in two phases, a liquid phase denoted by subscript $\mathrm{L}$ and a vapor phase denoted by subscript $\mathrm{V}$. In each phase the fluid is thermodynamically described by the corresponding Helmholtz free energy density $\varrho f_{\mathrm{L/V}}(\varrho)$. The fluid occupies a domain $\Omega \subset \mathbb{R}^d, \ d\in \mathbb{N}$.
Let $\varrho >0 $ be the density of the fluid, $\vec{v} \in \mathbb{R}^d$ the velocity and $\varphi \in [0,1]$ the phase field. Following \cite{Dreyer2014} we assume that the dynamics of the fluid is described by the isothermal compressible Navier--Stokes--Allen--Cahn system.
\begin{align}
\partial_t \varrho + \operatorname{div}(\varrho \mathbf{v}) &= 0, \label{eq:NSAC1}\\
\partial_t(\varrho \mathbf{v}) + \operatorname{div}(\varrho \mathbf{v}\otimes \mathbf{v}+{p\mathbf{I}})&= \operatorname{div}(\mathbf{S}) - \gamma \operatorname{div}(\nabla \varphi\otimes \nabla \varphi) \ \text{ in } \Omega \times (0,T), \label{eq:NSAC2}\\
\partial_t(\varrho \varphi) + \operatorname{div}(\varrho \varphi \mathbf{v}) &= -\eta \mu. \label{eq:NSAC3}  
\end{align}
The Helmholtz free energy density $\varrho f$ is defined as 
\begin{align}
\varrho f(\varrho,\varphi,\nabla\varphi) &= h(\varphi)\varrho f_\mathrm{L}(\varrho)+(1-h(\varphi)) \varrho f_\mathrm{V}(\varrho) + \frac{1}{\gamma}W(\varphi) + \frac{\gamma}{2}|\nabla\varphi|^2 \label{eq:rhof} \\
&=: \varrho \psi(\varphi,\varrho) + \frac{1}{\gamma}W(\varphi) + \frac{\gamma}{2} |\nabla \varphi|^2.
\end{align}
It consists of the interpolated free energy densities $\varrho f_{\mathrm{L/V}}$ of the pure liquid and vapor phases with the nonlinear interpolation function 
\begin{align}\label{eq:def_h}
h(\varphi) = 3 \varphi^2 - 2 \varphi^3,
\end{align}
and a mixing energy \cite{Cahn1958} using the double well potential $W(\varphi)=\varphi^2(1-\varphi)^2$.

The hydrodynamic pressure $p$ is determined through the Helmholtz free energy $\varrho f$ by the thermodynamic relation
\begin{equation}\label{eq:pdef}
p=p(\varrho,\varphi) = -\varrho f(\varrho,\varphi)+\varrho \frac{\partial (\varrho f)}{\partial \varrho}(\varrho, \varphi).
\end{equation}
We define the generalized chemical potential 
\begin{equation}
\mu = \frac{1}{\gamma}W'(\varphi)+ \frac{\partial (\varrho\psi)}{\partial \varphi}-\gamma \Delta \varphi,
\end{equation}
which steers the phase field variable into equilibrium.
Additionally, we denote by $\eta>0$ the (artificial) mobility.

The dissipative viscous part of the stress tensor reads as $\mathbf{S}=\mathbf{S}(\varphi,\nabla\mathbf{v})= \nu(\varphi) (\nabla \mathbf{v}+ \nabla \mathbf{v}^\top - \operatorname{div}(\mathbf{v})\mathbf{I})$ with an interpolation of the viscosities $\nu_{\mathrm{L/V}}$ of the pure phases $\nu(\varphi) = h(\varphi)\nu_\mathrm{L} + (1-h(\varphi))\nu_\mathrm{V} > 0$.

The total energy of the system \eqref{eq:NSAC1}-\eqref{eq:NSAC3} at time $t$ is defined as
\begin{align}
\label{eq:etot}
E(t) &\mathrel{\vcenter{\baselineskip0.5ex \lineskiplimit0pt
                     \hbox{\scriptsize.}\hbox{\scriptsize.}}}%
                     = E_\mathrm{free}(t) + E_\mathrm{kin}(t) \nonumber\\ 
                     &= \int_\Omega \varrho(\vec{x},t) f(\varrho(\vec{x},t),\varphi(\vec{x},t),\nabla \varphi(\vec{x},t)) + \frac{1}{2} \varrho(\vec{x},t) |\mathbf{v}(\vec{x},t)|^2 \operatorname{~ d\!} \mathbf{x}.
\end{align}

\begin{remark} \
\begin{enumerate}
\item 
The phase field $\varphi$ is in general an artificial variable, however in this case it can be viewed as a mass fraction $\varphi = \frac{m_\mathrm{V}}{m},$ with the mass $m_\mathrm{V}$ of the vapor constituent and the total mass $m$ of the fluid.
\item  
The special form of the nonlinear interpolation function $h$ with $h'(0) = h'(1) \neq 0$ guarantees that $\eqref{eq:NSAC1}-\eqref{eq:NSAC3}$ allows for physical meaningfull equilibria. This can be easily seen by considering a static single-phase equilibrium $\vec{v} = \boldsymbol{0}, \varphi \equiv 0$. If $h'(0)\neq 0$ then the right hand side of the phase field equation \eqref{eq:NSAC3} does not vanish.
\end{enumerate}
\end{remark}
Assuming an impermeable wall, the velocity must satisfy the boundary condition
\begin{equation}
\mathbf{v}\cdot \mathbf{n} = 0 \quad \text{ on $\partial \Omega$}. \label{eq:bc1}
\end{equation}
Additionally, the system is endowed with initial conditions
\begin{align}
\label{eq:IC}
\varrho = \varrho_0, \quad \vec{v} = \vec{v}_0, \quad \varphi = \varphi_0 \quad \text{ on } \Omega \times \{0\},
\end{align}
using suitable functions $(\varrho_0,\vec{v}_0, \varphi_0)\colon \Omega \to \mathbb{R}^+ \times \mathbb{R}^d \times [0,1]$.

However, in order to close the system \eqref{eq:bc1} does not suffice. In the following section we derive a complete set of boundary conditions that allow for  moving contact lines (MCL).

\subsection{Boundary Conditions}
\label{sec:bc}
\begin{figure}[h]
\centering\includegraphics[width=.649\textwidth]{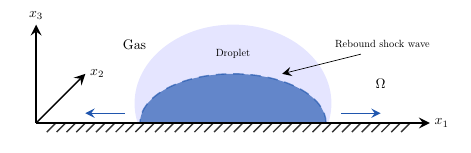}
\caption{Sketch of a compressible droplet impingement on a flat wall with moving contact line.}
\label{fig:MCL}
\end{figure}

The system \eqref{eq:NSAC1}-\eqref{eq:NSAC3} needs to be complemented with initial and boundary conditions. We are interested in MCL problems. With a sharp interface point of view, the contact line is the intersection of the liquid-vapor interface with the solid wall. The requirement of a contact line moving along the wall renders the derivation of boundary conditions nontrivial. Figure \ref{fig:MCL} depicts a sketch of a compressible droplet impact scenario with the rebound shock wave dynamics and a moving contact line.
We derive appropriate boundary conditions to handle MCL problems with the phase field system \eqref{eq:NSAC1}-\eqref{eq:NSAC3} in this section.

For the incompressible case so called \emph{general Navier boundary conditions} (GNBC) have been derived \cite{Qian2003,QIAN2006}. Motivated by these works we extend GNBC to the compressible case.

Because phase field modelling goes well with energy principles we add a wall free energy term $\int_{\partial \Omega} g(\varphi) \operatorname{~ d\!} s$ to the total energy $E$ from \eqref{eq:etot} and obtain 
\begin{align}\label{eq:total_energy}
E_{\mathrm{tot}}(t) &= E(t) + E_{\mathrm{wall}}(t)  \nonumber\\
&=  \int_\Omega \varrho(t) f(\varrho(t),\varphi(t),\nabla \varphi(t)) + \frac{1}{2} \varrho(t) |\mathbf{v}(t)|^2 \operatorname{~ d\!} \mathbf{x} + \int_{\partial \Omega} g(\varphi(t)) \operatorname{~ d\!} s.
\end{align}

\begin{figure}[h]
\centering\includegraphics[width=.6\textwidth]{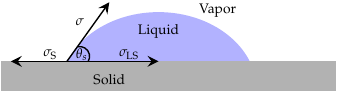}
\caption{Illustration of Young's equation $\sigma \cos(\theta_s)= \sigma_\mathrm{S}-\sigma_\mathrm{LS}.$}
\label{fig:young}
\end{figure}

Here $g(\varphi)$ is the interfacial free energy per unit area at the fluid-solid boundary depending only on the local composition \cite{QIAN2006}. The specific choice for $g$ is motivated by Young's equation.
With a sharp interface point of view we have
\begin{align}\label{eq:young}
\sigma \cos(\theta_s)= \sigma_\mathrm{S}-\sigma_\mathrm{LS},
\end{align}
with the surface free energy $\sigma$ of the liquid, the static contact angle $\theta_\mathrm{s}$, surface free energy $\sigma_\mathrm{S}$ of the solid, and interfacial free energy $\sigma_{\mathrm{LS}}$ between liquid and solid, see Figure \ref{fig:young}. We prescribe the difference in energy for $g$, i.e.
\begin{align}
\sigma_\mathrm{S}-\sigma_\mathrm{LS}=g(0)-g(1).
\end{align}

Then, we choose a smooth interpolation between the values $\pm\frac{\Delta g}{2} = \pm \frac{g(1)-g(0)}{2}$. However, it was shown in \cite{Qian2003} that the choice of the kind interpolation has no large impact. Hence, for reasons of consistency we use $h$ as interpolation function.
With \eqref{eq:young} we obtain  
\begin{equation}
g(\varphi) \mathrel{\vcenter{\baselineskip0.5ex \lineskiplimit0pt
                     \hbox{\scriptsize.}\hbox{\scriptsize.}}}%
                     = -\sigma
\cos(\theta_s) \left(h(\varphi)-\frac{1}{2}\right).
\end{equation}                                          
A variation $\delta \varphi$ of $\varphi$ leads to a variation $\delta E_\mathrm{tot}$ of the total energy \eqref{eq:total_energy}, that is
\begin{align*}
\delta E_\mathrm{tot} = \int_\Omega \mu \delta \varphi \operatorname{~ d\!} \vec{x} - \int_{\partial \Omega} L(\varphi)\frac{\partial\varphi}{\partial_{\boldsymbol{\tau}}} \delta\varphi_{\boldsymbol{\tau}}.
\end{align*}
Here,
\[L(\varphi) \coloneqq \gamma \frac{\partial \varphi}{\partial \mathbf{n}}+g'(\varphi)\] 
can be interpreted as uncompensated Young stress \cite{Qian2003}. The boundary tangential vector is denoted by $\boldsymbol{\tau}$  and $\vec{n}$ denotes the outer normal. Thus, $L(\varphi)=0$ is the Euler--Lagrange equation at the fluid-solid boundary for minimizing the total energy \eqref{eq:total_energy} with respect to the phase field variable. We assume a boundary relaxation dynamics for $\varphi$ given by
\begin{equation}
\partial_t \varphi + \mathbf{v}\cdot \nabla_{\boldsymbol{\tau}} \varphi = -\frac{\alpha}{\varrho} L(\varphi),
\end{equation}
with a relaxation parameter $\alpha>0$. Here $\nabla_{\boldsymbol{\tau}} \mathrel{\vcenter{\baselineskip0.5ex \lineskiplimit0pt
                     \hbox{\scriptsize.}\hbox{\scriptsize.}}}%
                     = \nabla-(\mathbf{n}\cdot\nabla)\mathbf{n}$ is the gradient along the tangential direction. 
                     Since $\mathbf{v} \cdot \mathbf{n} = 0$, we have $\mathbf{v}\cdot \nabla_{\boldsymbol{\tau}}\varphi = v_{\boldsymbol{\tau}}\frac{\partial \varphi}{\partial \boldsymbol{\tau}}$,
and finally we obtain
\begin{align}
\partial_t \varphi + v_{\boldsymbol{\tau}}\frac{\partial \varphi}{\partial \boldsymbol{\tau}} &= -\frac{\alpha}{\varrho} L(\varphi) \quad \text{ on $\partial \Omega$}.\label{eq:bc3}
\end{align}

In order to complete the derivation of the GNBC we incorporate a slip velocity boundary condition. In single phase models the slip velocity is often taken proportional to the tangential viscous stress. However, in our case we also have to take the uncompensated Young stress into account. In \cite{Qian2003} it is shown from molecular dynamic simulations that the slip velocity should be taken proportional to the sum of the tangential viscous stress and the uncompensated Young stress.
Hence, with the slip length $\beta > 0$ we prescribe the boundary condition
\begin{equation}
\beta v_{\boldsymbol{\tau}} + \nu(\varphi) \frac{\partial v_{\boldsymbol{\tau}}}{\partial \mathbf{n}} - L(\varphi)
\frac{\partial \varphi}{\partial \boldsymbol{\tau}} =0 \quad \text{ on $\partial \Omega$}. \label{eq:bc2}
\end{equation}

Away from the interface the last term in \eqref{eq:bc2} drops out and we have the classical Navier-slip condition but in the interface region the additional term acts and allows for correct contact line movement.

In summary we obtain the following GNBC for the MCL problem
\begin{align}
\mathbf{v}\cdot \mathbf{n} &= 0, \label{eq:gnbc1} \\
\beta v_{\boldsymbol{\tau}} + \nu(\varphi) \frac{\partial v_{\boldsymbol{\tau}}}{\partial \mathbf{n}} - L(\varphi)
\frac{\partial \varphi}{\partial \boldsymbol{\tau}} &=0, \hspace*{6ex} \text{ on $\partial \Omega$.} \label{eq:gnbc2} \\
\partial_t \varphi + v_{\boldsymbol{\tau}}\frac{\partial \varphi}{\partial \boldsymbol{\tau}} &= -\frac{\alpha}{\varrho} L(\varphi)\label{eq:gnbc3}
\end{align}

The GNBC \eqref{eq:gnbc1}, \eqref{eq:gnbc2}, \eqref{eq:gnbc3} contain certain subcases. For $\alpha \to \infty$ we obtain the static contact angle boundary condition and with $\beta \to \infty$ we end up with no-slip boundray conditions.


\subsection{Energy Inequality}
\label{sec:energy_ineq}
For isothermal models thermodynamical consistency means to verify that solutions of the problem at hand admit an energy inequality. Precisely, we have for the system \eqref{eq:NSAC1}-\eqref{eq:NSAC3} the following result.
\begin{thm}[Energy inequality]\label{thm:energy_ineq}
Let $(\varrho,\mathbf{v},\varphi)$ with values in $(0,\infty)\times \mathbb{R}^d \times [0,1]$ be a classical solution of \eqref{eq:NSAC1}-\eqref{eq:NSAC3} in $(0,T) \times \Omega$ satisfying the boundary conditions \eqref{eq:gnbc1} - \eqref{eq:gnbc3} on $(0,T) \times \partial \Omega$. Then for all $t \in (0,T)$ the following energy inequality holds:
\begin{flalign} \label{eq:energy_ineq}
&\frac{\operatorname{d}}{\operatorname{~ d\!} t} E_\mathrm{tot}(t) = \frac{\operatorname{d}}{\operatorname{~ d\!} t} (E_\mathrm{free}(t) + E_\mathrm{kin}(t) +E_\mathrm{wall}(t)) &  \nonumber\\
=&\frac{\operatorname{d}}{\operatorname{~ d\!} t} \left(\int_\Omega \varrho f(\varrho,\mathbf{v},\varphi,\nabla \varphi) + \frac{1}{2} \varrho |\mathbf{v}|^2 \operatorname{~ d\!} \mathbf{x} + \int_{\partial \Omega} g(\varphi) \operatorname{~ d\!} s\right)  & \nonumber \\ 
=&-\int_\Omega \frac{\eta}{\varrho} \mu^2 \operatorname{~ d\!} \mathbf{x} - \int_\Omega \mathbf{S} \colon \nabla \mathbf{v} \operatorname{~ d\!} \mathbf{x}  \nonumber\\
&- \int_{\partial \Omega} \beta |v_{\boldsymbol{\tau}}|^2 \operatorname{~ d\!} s - \int_{\partial \Omega} \frac{\alpha}{\varrho} |L(\varphi)|^2  \operatorname{~ d\!} s \leq 0.
\end{flalign}
\end{thm}
As expected the energy inequality renders phase transition, viscosity, wall slip, and composition relaxation at the solid interface to be drivers of energy with respect to entropy dissipation.
\begin{proof}
In a straightforward way we compute:

\begin{align*}
\frac{\operatorname{~ d\!}}{\operatorname{~ d\!}t} E_\mathrm{tot}(t) =& \frac{\operatorname{~ d\!}}{\operatorname{~ d\!}t} \left(\int_\Omega \varrho f(\varrho,\varphi,\nabla\varphi) + \frac{1}{2}\varrho |\vec{v}|^2 \operatorname{~ d\!}\vec{x} + \int_{\partial \Omega} g(\varphi) \operatorname{~ d\!} s \right) \\
=& \frac{\operatorname{~ d\!}}{\operatorname{~ d\!}t} \left(\int_\Omega\frac{1}{\gamma} W(\varphi) + \varrho\psi(\varrho,\varphi) + \frac{\gamma}{2} |\nabla\varphi|^2  + \frac{1}{2}\varrho |\vec{v}|^2 \operatorname{~ d\!} \vec{x}+  \int_{\partial \Omega} g(\varphi) \operatorname{~ d\!} s \right)\\
=& \int_\Omega \varphi_t \left(\frac{1}{\gamma}W'(\varphi)+\frac{\partial (\varrho \psi)}{\partial \varphi}-\gamma\Delta \varphi\right) + \varrho_t \left(\frac{\partial (\varrho \psi)}{\partial \varrho} - \frac{1}{2} |\vec{v}|^2\right) + (\varrho\vec{v})_t \cdot \vec{v} \operatorname{~ d\!}\vec{x}   \\
&+ \int_{\partial \Omega} \varphi_t(g'(\varphi) +  \gamma \nabla \varphi \cdot \vec{n})  \operatorname{~ d\!} s. \\
\intertext{Now we use \eqref{eq:NSAC1}-\eqref{eq:NSAC3} to replace the time derivatives in the volume integrals. Using \eqref{eq:pdef} we obtain after basic algebraic manipulations}
\frac{\operatorname{~ d\!}}{\operatorname{~ d\!}t} E_\mathrm{tot}(t) =&  - \int_\Omega \operatorname{div}(\varrho\vec{v})\left(\frac{\partial (\varrho\psi)}{\partial\varrho}-\frac{1}{2}|\vec{v}|^2\right) + \operatorname{div}(\varrho \vec{v}\otimes\vec{v})\cdot\vec{v} \operatorname{~ d\!} \vec{x} - \int_\Omega \frac{\eta}{\varrho} \mu^2  \operatorname{~ d\!} \vec{x}  \\
&- \int_\Omega \vec{v}\cdot \varrho \nabla\left(\frac{\partial (\varrho \psi)}{\partial \varrho}\right) - \operatorname{div}(\vec{S})\cdot\vec{v} \operatorname{~ d\!} \vec{x}
+ \int_{\partial \Omega} \varphi_t L(\varphi) \operatorname{~ d\!} s. \\ \intertext{We integrate by parts and have}
\frac{\operatorname{~ d\!}}{\operatorname{~ d\!}t} E_\mathrm{tot}(t) =& - \int_\Omega \frac{\eta}{\varrho} \mu^2  \operatorname{~ d\!} \vec{x} - \int_\Omega \vec{S}\colon \nabla \vec{v} \operatorname{~ d\!} \vec{x}+ \int_{\partial \Omega} \varphi_t L(\varphi)  \operatorname{~ d\!} s \\
&+ \int_{\partial \Omega} \vec{S}\vec{v}\cdot\vec{n} - \varrho\vec{v}\left(\frac{\partial (\varrho \psi)}{\partial \varrho}+\frac{1}{2}|\vec{v}|^2\right)\cdot\vec{n}\operatorname{~ d\!} s.  \\ \intertext{With the boundary conditions \eqref{eq:gnbc1}-\eqref{eq:gnbc3} we finally obtain}
\frac{\operatorname{~ d\!}}{\operatorname{~ d\!}t} E_\mathrm{tot}(t) =& - \int_\Omega \frac{\eta}{\varrho} \mu^2  \operatorname{~ d\!} \vec{x} - \int_\Omega \vec{S}\colon \nabla \vec{v} \operatorname{~ d\!} \vec{x} - \int_{\partial \Omega} \beta |v_{\boldsymbol{\tau}}|^2 \operatorname{~ d\!} s - \int_{\partial \Omega} \frac{\alpha}{\varrho} |L(\varphi)|^2  \operatorname{~ d\!} s.
\end{align*}
This concludes the proof.
\end{proof}

\subsection{Surface Tension}
\label{sec:surface_tens}
There are different interpretations of surface tension. It can be either viewed as a force acting in tangential direction of the interface or as excess energy stored in the interface \cite{Jamet2002}.
In line with our energy-based derivation we consider a planar equilibrium profile and integrate the excess free energy density over this profile.
We assume that static equilibrium conditions hold, i.e. $\vec{v} = \boldsymbol{0}.$
The planar profile is assumed to be parallel to the $x$-axis and density, velocity and phase field are independent from $t, y,$ and $z$. 
Then the equilibrium is governed by the solution of the following boundary value problem on the real line. 

Find $\varrho=\varrho(x), \varphi=\varphi(x)$ such that
\begin{align}
\left(-\varrho\psi-\frac{1}{\gamma}W(\varphi) - \frac{\gamma}{2}\varphi_x^2 +\varrho \frac{\partial (\varrho \psi)}{\partial\varrho}\right)_x &= -{\gamma(\varphi_x^2)}_x, \label{eq:eq1} \\
\frac{1}{\gamma} W'(\varphi) + \frac{\partial (\varrho \psi)}{\partial \varphi} - \gamma \varphi_{xx} &= 0,\label{eq:eq2}
\end{align}
and
\begin{equation}
\varrho(\pm\infty) = \varrho_\mathrm{V/L}, \quad \varphi(-\infty) = 0 ,  \quad \varphi(\infty) = 1, \quad \varphi_x(\pm\infty)=0. \label{eq:eqbc}
\end{equation}

Multiplying \eqref{eq:eq2} with $\varphi_x$ and substracting from \eqref{eq:eq1} yields
\begin{equation}
\frac{\partial (\varrho \psi)}{\partial \varrho} = const.  
\label{eq:muconst}
\end{equation}
Multiplying \eqref{eq:eq2} with $\varphi_x$, integrating from $-\infty$ to some $x\in \mathbb{R}$ using \eqref{eq:eq1} and \eqref{eq:eqbc} leads to

\begin{equation}
\frac{1}{\gamma} W(\varphi(x)) + \varrho(x)\psi(\varrho(x),\varphi(x)) - \varrho_\mathrm{V}(x)\psi(\varrho_\mathrm{V}(x),0) = \frac{\gamma}{2}\varphi_x^2(x).
\label{eq:surfacetenshelp}
\end{equation}
From \eqref{eq:surfacetenshelp} we obtain for $x\to \infty$
\begin{equation}
\varrho_\mathrm{L}\psi(\varrho_\mathrm{L},1) = \varrho_\mathrm{V}\psi(\varrho_\mathrm{V},0) \eqqcolon \overline{\varrho\psi}.
\label{eq:surfacetenshelp1}
\end{equation}

As mentioned before, surface tension can be defined by means of excess free energy. Roughly speaking an excess quantity is the difference of the quanity in the considered system and in a (sharp interface) reference system where the bulk values are maintained up to a dividing interface.
The interface position $x_0$ is determined by vanishing excess density.
%

In summary we define surface tension $\sigma$ via the relationship
\begin{align}
\sigma =& \int_{-\infty}^{x_0} \varrho f(\varrho,0,\varphi,\varphi_x) - \varrho_\mathrm{V}\psi(\varrho_\mathrm{V},0) \operatorname{~ d\!} x \nonumber  \\
&+\int_{x_0}^\infty \varrho f(\varrho,0,\varphi,\varphi_x) - \varrho_\mathrm{L}\psi(\varrho_\mathrm{L},1) \operatorname{~ d\!} x, \\
\intertext{where $(\varrho,\varphi)$ is a solution of \eqref{eq:eq1}-\eqref{eq:eqbc}. Using \eqref{eq:surfacetenshelp} we have}
\sigma =& \int_{-\infty}^{x_0} \gamma \varphi_x^2  \operatorname{~ d\!} x + \int_{x_0}^\infty \gamma \varphi_x^2 + (\varrho_\mathrm{V}\psi(\varrho_\mathrm{V},0)-\varrho_\mathrm{L}\psi(\varrho_\mathrm{L},1)) \operatorname{~ d\!} x. \\ 
\intertext{With \eqref{eq:surfacetenshelp1} it follows}
\sigma =& \int_{-\infty}^{\infty} \gamma \varphi_x^2  \operatorname{~ d\!} x  = \sqrt{2} \int_{\varphi_\mathrm{V}}^{\varphi_\mathrm{L}} \sqrt{W(\varphi)+\gamma(\varrho\psi(\varrho(\varphi),\varphi)- \overline{\varrho\psi})} \operatorname{~ d\!} \varphi.
\end{align}

In the last step we used the transformation from $x$ to $\varphi$ integration. This is possible since $\varrho$ can be written in dependence on $\varphi$:  Assuming convex free energies $\varrho f_\mathrm{L/V}$ in \eqref{eq:rhof}, we have convex $\varrho\psi$ in $\varrho$ and from \eqref{eq:muconst} follows with the implicit function theorem $\varrho = \varrho(\varphi)$.

One can see that the surface tension is mainly dictated by the double well potential $W(\varphi)$. There is a contribution due to the equations of state of the different phases, however in the sharp interface limit, i.e. $\gamma \to 0$ this contribution vanishes. This is a difference to (quasi-)incompressible models like \cite{Lowengrub1998}. There is no contribution due to the equation of states and the surface tension is purely determined by the double well function. Of course surface tension is a material parameter and given by physics dependening on the fluids and walls considered. Therefore, in simulations the double well should be scaled accordingly to yield the correct surface tension.

\section{Numerical Experiments}
\label{sec:num_exp}
The phase field system \eqref{eq:NSAC1}-\eqref{eq:NSAC3} is of mixed hyperbolic-parabolic type. This complicates the derivation of discretization methods. An appropriate choice are discretizations based on discontinuous Galerkin methods. In fact even versions which reproduce the energy dissipation precisely are available \cite{Giesselmann2015a,Repossi2017,Kraenkel2018}. The key idea behind those schemes is to achieve stabilization through the exact approximation of the energy, that means the energy inequality \eqref{eq:energy_ineq} should be fullfilled exactly on the discrete level without introducing numerical dissipation. This helps to prevent increase of energy and possibly associated spurious currents. Additionally, the schemes are designed such that they preserve the total mass. Motivated by \cite{Giesselmann2015a,Kraenkel2018}  we derived such a scheme for the system \eqref{eq:NSAC1}-\eqref{eq:NSAC3},
for details we refer to \cite{Ostrowski2018}.
In the following we present three numerical simulations using this scheme.

\subsection{Choice of Parameters}
For the equations of state in the bulk phases, we choose stiffened gas equations
\begin{align*}
\varrho f_\mathrm{L/V}(\varrho) = \alpha_\mathrm{L/V} \varrho \ln(\varrho) + (\beta_\mathrm{L/V}-\alpha_\mathrm{L/V})\varrho + \gamma_\mathrm{L/V},
\end{align*}
with parameters $\alpha_\mathrm{L/V} > 0,  \beta_\mathrm{L/V} \in \mathbb{R}, \gamma_\mathrm{L/V} \in \mathbb{R}$.
In order to avoid prefering on of the phases, we choose the minima of the two free energies to be at the same height.

Due to surface tension the density inside a liquid droplet is slightly higher than the value which minimizes $\varrho f_\mathrm{L}$. The value of the surrounding vapor is slightly lower than the minimizer of $\varrho f_\mathrm{V}$. We choose the initial density profile accordingly. For the bulk viscosities we set $\nu_\text{L} = 0.0125$ and $\nu_\text{V} = 0.00125$. If not stated otherwise, the capillary parameter is taken $\gamma=5\cdot 10^{-4}$ and the mobility $\eta=10$. The polynomial order of the DG polynomials is $2$.

\subsection{Merging Droplets}\label{sec:ex1}

\begin{figure}[h]
\centering\includegraphics[width=.9\textwidth]{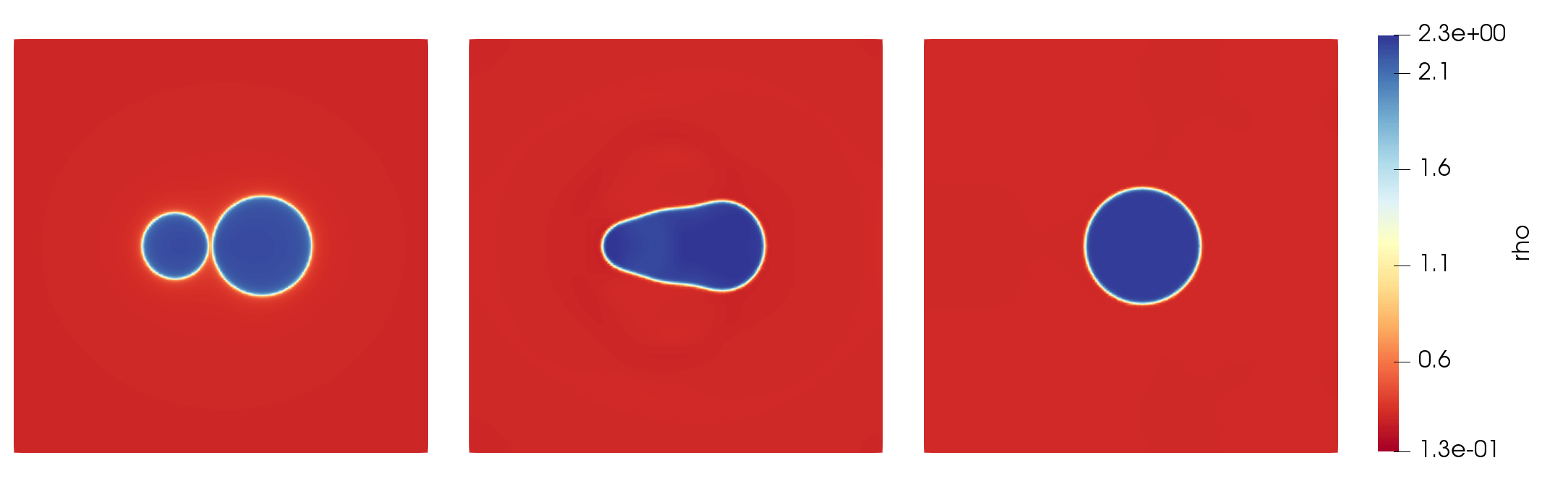}
\caption{Merging droplets. Density $\varrho$ at times $t=0$, $t=0.2$, and $t=2$ for $\eta = 10.$}
\label{fig:merging}
\end{figure}

In order to illustrate that phase field models are able to handle topological changes, we consider the example of two merging droplets. Initially we have no velocity field, $\vec{v}_0 = \boldsymbol{0}$ and look at two kissing droplets. The computational domain is $[0,1]\times[0,1]$. The droplets are located at $(0.39,0.5)$ and $(0.6,0.5)$ with radii $0.08$ and $0.12$. The parameters for the equations of states are $\alpha_\mathrm{L} = 5 ,\beta_\mathrm{L}= -4, \gamma_\mathrm{L}= 11,\alpha_\mathrm{V} = 1.5 ,\beta_\mathrm{V}= 1.8, \gamma_\mathrm{V}=0.324$. The inital density profile is smeared out with value $\varrho_\mathrm{L}=2.23$ inside and $\varrho_\mathrm{V} = 0.3$ outside the droplet. As expected the droplets merge into one larger droplet. This evolution with $\eta = 10$ is depicted in Figure \ref{fig:merging}.

\begin{SCfigure}
\centering\includegraphics[height=.4\textwidth]{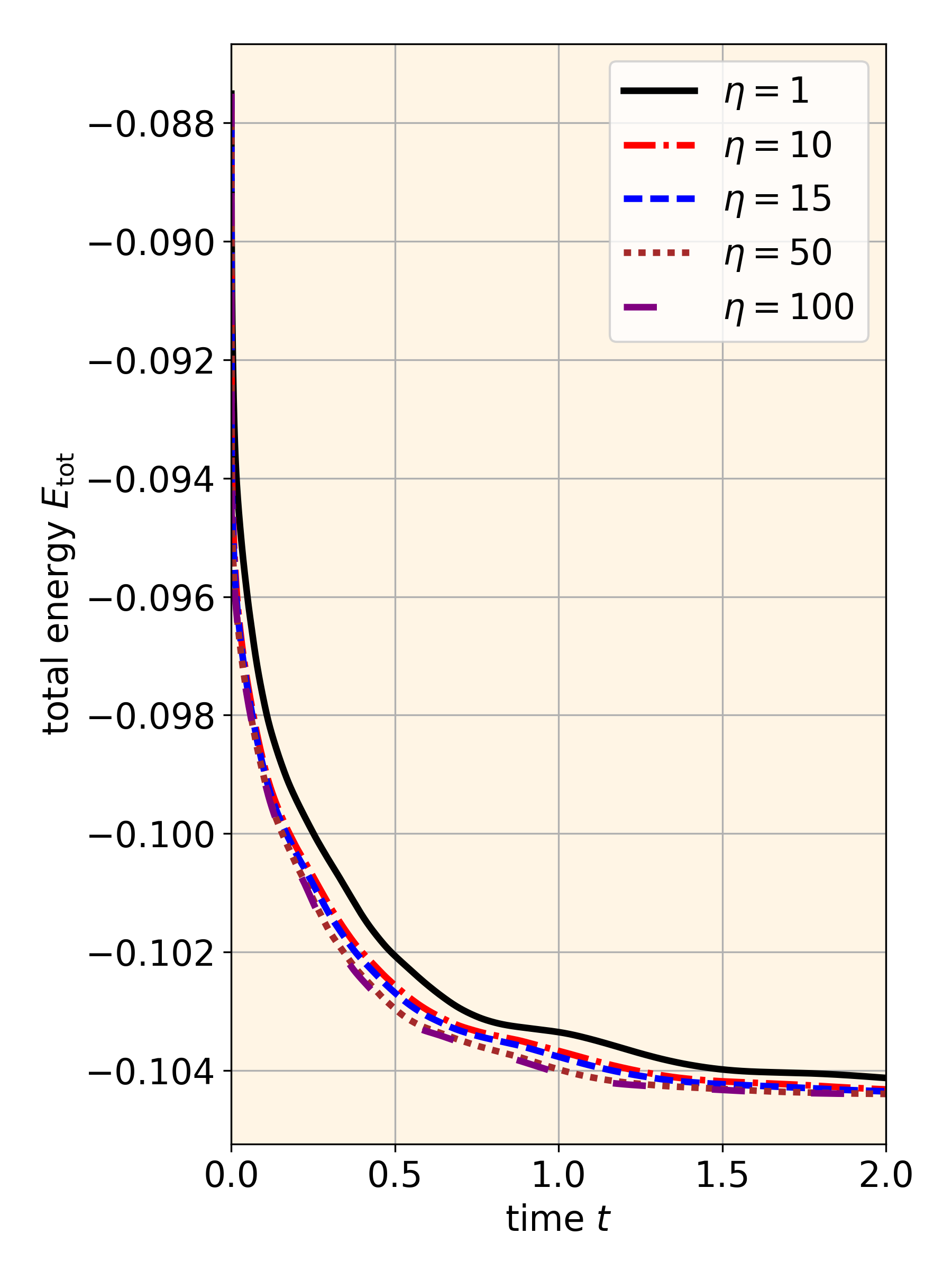}
\caption{Total energy $E_\mathrm{tot}$ over time for droplet merging simulation with different values for the mobility $\eta$.}
\label{fig:mergingenergy}
\end{SCfigure}

We can observe that the model handles topological changes easily.  However, the dynamics of the phase field relaxation are determined by the mobility $\eta$ which needs to be chosen according to the problem. This is illustrated in Figure \ref{fig:mergingenergy}, where the total energy over time for different values of the mobility $\eta$ is plotted.

The numerical scheme is designed to mimic the energy inequality \eqref{eq:energy_ineq} on the discrete level. The discrete energy decreases, as expected from \eqref{eq:energy_ineq} the higher the value of $\eta$, the higher the energy dissipation.

\subsection{Contact Angle}
\label{sec:ex2}
\begin{figure}[h]
\centering\includegraphics[width=.9\textwidth]{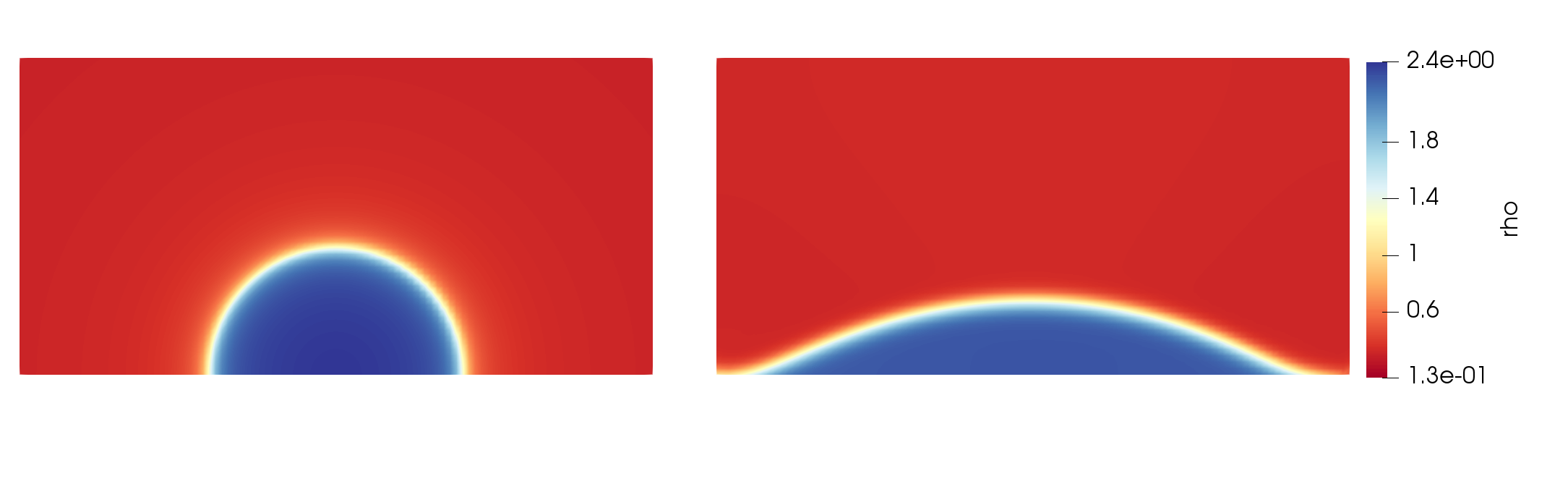}
\caption{Wetting of smooth wall with (GNBC) boundary conditions for the static limit $\alpha\to\infty$ and contact angle $\theta_s=0.1\pi$. Density $\varrho$ at $t=0$ and $t=0.9$.}
\label{fig:wetting}
\end{figure}

In this example we adress droplet wall interactions. We consider the case of static contact angle. This means we let the relaxation parameter $\alpha$ in \eqref{eq:bc3} tend to infinity.
In the limit we obtain the static contact angle boundary conditions:
%

We set the static contact angle $\theta_s = 0.1\pi \approx 18^\circ$. The computational domain, density values and EOS parameters are like in Section \ref{sec:ex1}. As initial condition we use a droplet sitting on a flat surface with a contact angle of $90^\circ$. The droplet position is $(0.5,0)$ with radius $0.2$. Since the initial condition is far away from equilibrium we have dynamics on the wall-boundary towards the equilibrium configuration. Thus, we can observe a wetting dynamic, see Figure \ref{fig:wetting}. 

The wall contribution leads to a large force on the boundary, which renders the system stiff. Although we have an implicit scheme we increased the interface width to be able to handle the boundary terms. Hence, we chose in this simulation $\gamma = 10^{-2}$.

\subsection{Droplet Impingement}
\label{sec:ex3}
\begin{figure}[h!]
\centering\includegraphics[width=.86\textwidth]{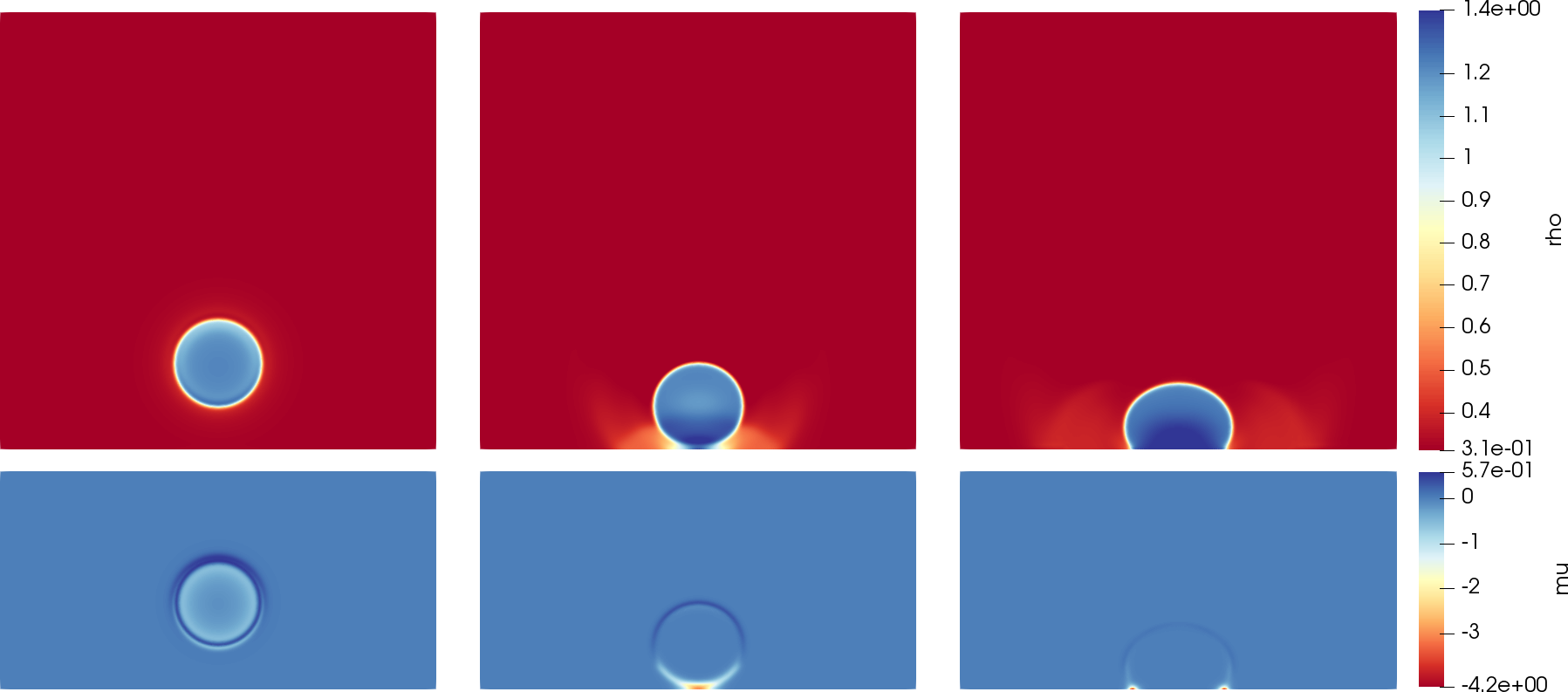}
%
\caption{
Droplet impact simulation. Density $\rho$ and chemical potential $\mu$ at times $t=0.005, t=0.13,t=0.21$.
}
\label{fig:impact}
\end{figure}

With this example we consider droplet impingement. The computational domain is the same as in Section \ref{sec:ex1}. As initial condition we use a droplet at $(0.5,0.2)$ with radius $0.1$.
The parameters for the equations of states are $\alpha_\mathrm{L} = 5 ,\beta_\mathrm{L}= -0.8, \gamma_\mathrm{L}= 5.5,\alpha_\mathrm{V} = 1.5 ,\beta_\mathrm{V}= 1.8, \gamma_\mathrm{V}=0.084$. The inital density profile is smeared out with value $\varrho_\mathrm{L}=1.2$ inside and $\varrho_\mathrm{V} = 0.3$ outside the droplet.
In contrast to sharp interface models based on the Navier--Stokes equations, phase field models can still have contact line movement even if no-slip conditions are used. This is due to the fact that the contact line is regularized and the dynamics are driven by evolution in the phase field variable rather than advective transport. This can be seen in Figure \ref{fig:impact} where a droplet impact with noslip conditions is simulated. This is a special case of the GNBC, with $\alpha \to \infty$ and $\beta\to \infty$. 

It can be seen that the generalized chemical potential $\mu$ is low at the contact line which leads to fast dynamics in the phase field. This leads to a moving contact line. Additionally, we can see the (smeared out) shock waves in the vapor phase and also in the liquid phase where the shocks move faster due to a higher speed of sound in the liquid phase.

\section{Summary and Conclusions}
In this work we presented a phase field approach to model and simulate compressible droplet impingement scenarios. To be precise, we introduced a compressible Navier-Stokes-Allen-Cahn model in Section \ref{sec:NSAC}. We discussed modelling aspects, with emphasis on the energy-based derivation. We highlighted the connection of thermodynamic consistency with an energy inequality. Further, we proved in Theorem \ref{thm:energy_ineq} that solutions to the system fulfill this inequality. Surface tension can be interpreted as excess free energy. We quantified the amount of surface tension present in the model in Section \ref{sec:surface_tens}. Moving contact line problems need special attention with respect to boundary conditions. Hence, physical relevant boundary conditions were derived as Generalized Navier Boundary Conditions in Section \ref{sec:bc}. In Section \ref{sec:num_exp} numerical examples were given. In future work we implement the general, dynamic version of the GNBC to obtain jetting phenomena in the impact case.

\section*{Acknowledgments}
The authors kindly acknowledge the financial support of this work by the Deutsche Forschungsgemeinschaft (DFG)
in the frame of the International Research Training Group "Droplet Interaction Technologies" (DROPIT).

\begin{small}
\bibliographystyle{abbrv}
\bibliography{literature}
\end{small}

\end{document}